\documentclass[twocolumn,aps,prb,superscriptaddress,longbibliography,floatfix, 10pt]{revtex4-2}

\usepackage[utf8]{inputenc} 
\usepackage{graphicx}
\usepackage{amssymb, amsmath}
\usepackage[dvipsnames]{xcolor}
\usepackage{bm}
\usepackage{tabularx}
\newcommand {\spc} {\nobreak\hspace{.16667em}}
\newcommand {\spctwo} {\nobreak\hspace{.05em}}
\usepackage{xr-hyper}
\usepackage{hyperref}
\usepackage[capitalise]{cleveref}

\begin{document}
\title{Flat bands and unconventional superconductivity in a simple model of metal-organic frameworks}

\author{M. F. Ohlrich}
\email{miriamohlrich@gmail.com}
\affiliation{School of Mathematics and Physics, The University of Queensland, 4072, Australia}

\author{E. M. Makaresz}
\affiliation{School of Mathematics and Physics, The University of Queensland, 4072, Australia}

\author{H. L. Nourse}
\thanks{Present Address: School of Chemistry, University of Sydney, NSW 2006, Australia}
\affiliation{Quantum Information Science and Technology Unit, Okinawa Institute of Science and Technology Graduate University, Onna-son, Okinawa 904-0495, Japan}

\author{B. J. Powell}
\email{powell@physics.uq.edu.au}
\affiliation{School of Mathematics and Physics, The University of Queensland, 4072, Australia}

\begin{abstract}
    We show that interference-induced flat bands (IFBs) are commonplace in reticular materials. 
    For example, several metal-organic frameworks, including the superconductor Cu-BHT, form kagome lattices with metals at the vertices and ligands along the bonds. Similar bipartite motifs are common in reticular materials. A tight-binding model on this lattice yields partially occupied IFBs at half-filling with large gaps between them and all other bands. Long-range hopping induces curvature in the bands but leaves them flatter and more isolated than those in twisted bilayer graphene. The slave boson theory of the $t$-$J$ model on this lattice shows unconventional superconductivity. Thus, crystal engineering provides a lattice-driven route to flat bands with high electronic densities.
\end{abstract}

\maketitle

Reticular materials are  crystalline frameworks  built by linking molecular building blocks with strong (e.g., covalent, dative) bonds;  metal-organic frameworks (MOFs), coordination polymers (CPs) and covalent organic frameworks (COFs) provide prominent examples \cite{YaghiSciAdv}. Many correlated insulating states occur in reticular materials  \cite{Dinca-review,Deanna-review,COF-review,Thom,Bernard,HenrySummary,HenryHalf,HenryC3,Lowe2024}. Therefore, it has been argued that these systems may display unconventional superconductivity when driven out of their insulating phases, e.g., by doping  \cite{Merino}, pressure, or control of chemistry \cite{Cu-BHT-theo}. This hypothesis received strong support from the recent discovery of superconductivity in copper(II) benzenehexathiolate ({Cu-BHT}) \cite{CuBHT-ACIE}, and from the  evidence of unconventional superconductivity and strong electronic correlations in this material \cite{TakenakaT2021Scsi}.

Contemporaneously, the discovery of superconductivity and strongly correlated insulating states  in twisted bilayer graphene  (TBG) revived interest in flat band superconductivity \cite{CaoYuan2018Usim,CaoInsulator}.  In TBG, a group of isolated `nearly flat bands', with a total bandwidth on the order of $10$\spc meV, and band gaps of the same order on either side \cite{CaoYuan2018Usim,PathakShivesh2021Aatb}, cross the Fermi energy.
In a flat band, the kinetic energy of all  electrons is equal 
(hence, the density of states is enormous). Thus, interactions determine the behavior of the system, amplifying strongly correlated phenomena, such as the fractional quantum Hall effect, high temperature superconductivity, charge density waves, and metal-insulator transitions \cite{BalentsLeon2020Sasc,BernevigReview, TovmasyanMurad2016Etae, HeikkiläTeroT.2016FBaa, HaseI.2023Fbfi}. 
However, TBG has  extremely low electron densities, and many of the most interesting predictions for flat bands require high electron densities \cite{BernevigCatalogue,BalentsLeon2020Sasc}.  Thus, discovering bulk materials with flat bands is  an important goal.

Reticular materials offer the unique possibility of “crystal engineering” \cite{Robson1,Robson2}: realizing structures by design. Therefore, an important task for theory is to identify structures that promise the realization of new physical phenomena.
We show below that reticular materials are ideally suited for the study of flat band physics.

\begin{figure}
    \centering
    \includegraphics[width=.9\linewidth]{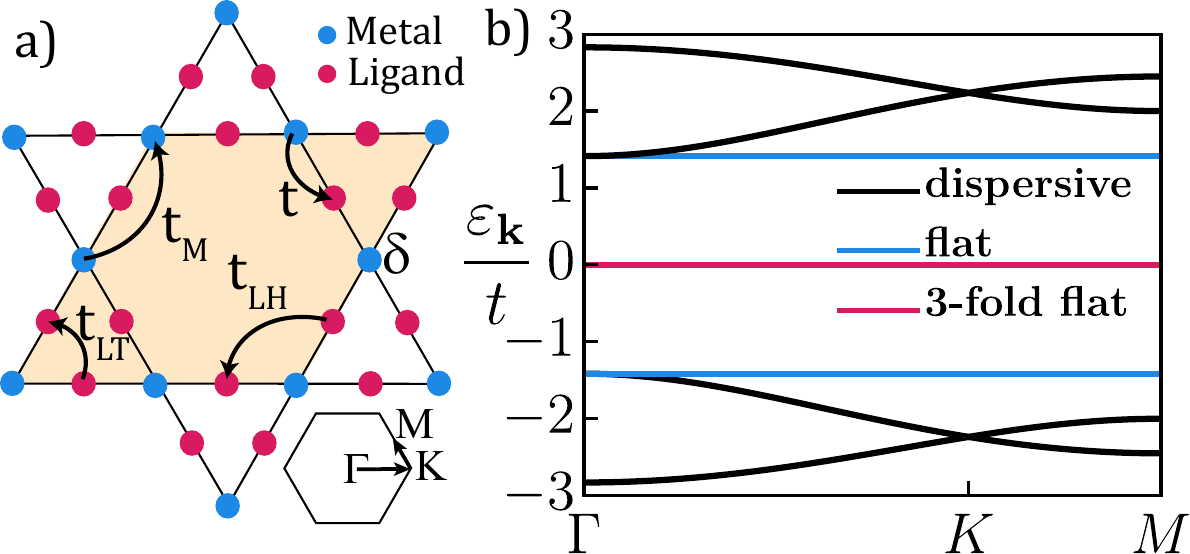}  
    \caption{Tight-binding model on the kagome-Lieb lattice. (a) Hopping integrals are marked, $\delta$ is the site energy of the metals, and the shaded area is the unit cell. The Bravais lattice is inset;  $\Gamma$, $M$ and $K$ label the high symmetry points. (b) Band structure of the nearest-neighbor model.}
    \label{fig:kagomelieb}
\end{figure}

\begin{figure}
    \centering
    \includegraphics[width=.9\linewidth]{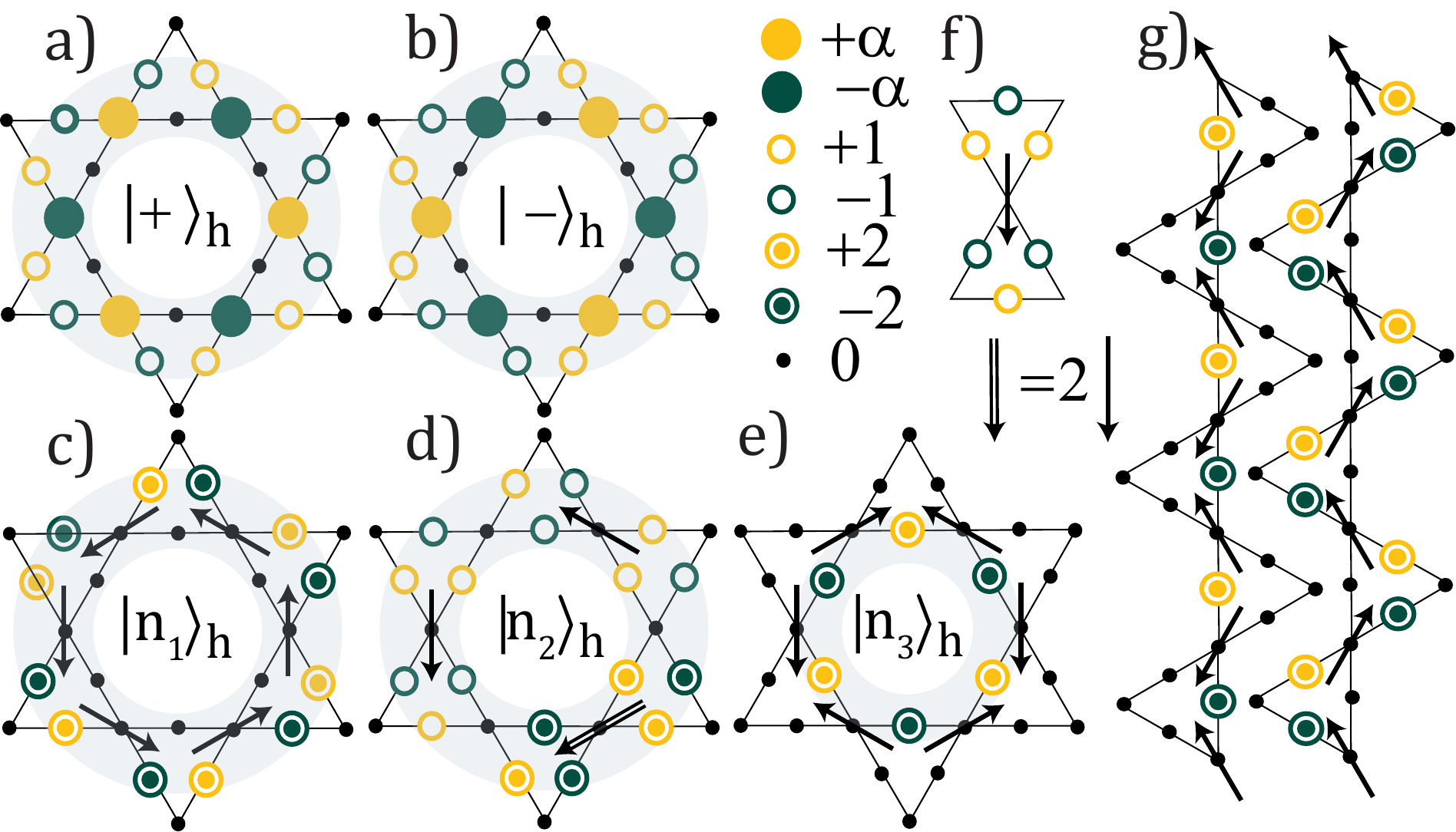}
    \caption{(Unnormalized) localized eigenstates for each of the interference-induced flat bands. Electrons cannot hop outside the loops  due to destructive interference. The (a) antibonding ($|+\rangle_h$) and  (b)  bonding ($|-\rangle_h$) kagome-like  eigenstates have energies  $E_{\pm} = [t_{\text{LT}}+ 2t_{\text{M}}+\delta \pm \sqrt{8t^2 + (2t_{\text{M}}-t_{\text{LT}}+\delta)^2}]/2$ and a relative amplitude on the metal sites of  $\alpha_{\pm} = \left( E_{\pm} - t_{\text{LT}} \right)/t$  (for $t_{\text{LH}}=0$).   The three  non-bonding  (between metals and ligands) eigenstates (c-e) can be constructed as superpositions of bow-tie states (f), as can noncontactable loops for PBC (g). The superpositions are indicated by the arrows which define the sign of the bow-tie wavefunctions centered on each metal site.}
    \label{fig:KLbalents}
\end{figure}

\begin{figure*}
    \centering
    \includegraphics[width=.9\linewidth]{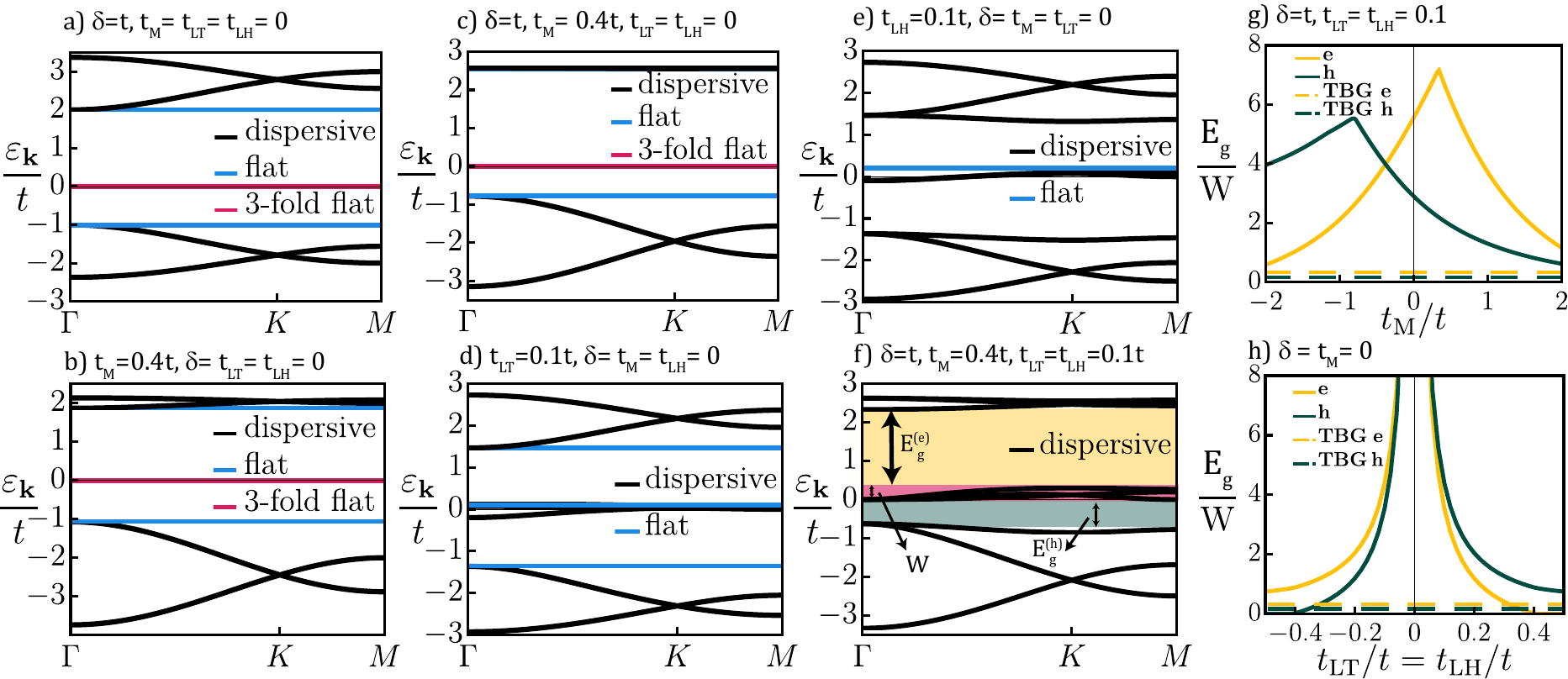}
    \caption{Evolution of the band structure of the kagome-Lieb lattice tight-binding model  with hopping beyond nearest-neighbors. Five bands remain exactly flat even when (a) metal-ligand orbital energy differences, $\delta$, (b) direct metal-metal hopping, $t_{\text{M}}$, or (c) both are included. (d) Ligand-ligand hopping across triangles,  $t_{\text{LT}}$,  leaves three exactly flat bands. (e) Ligand-ligand hopping across hexagons, $t_{\text{LH}}$, leaves one flat band. (f) Including all of these terms leaves no bands exactly flat, but the non-bonding bands remain gapped and  narrow. The bandwidth of the non-bonding bands, $W$, and the electron (hole) bandgap, $E_g^{(e)}$  ($E_g^{(h)}$), are labeled.  (g,h) Even for large $t_M, t_{LT}$ and $t_{LH}$, the non-bonding bands remain much flatter and more isolated (i.e., have larger $E_g/W$) than the nearly-flat bands in TBG [\onlinecite{PathakShivesh2021Aatb}].  
    } \label{fig:KLBS}  
\end{figure*}

\begin{figure}
    \includegraphics[width=0.9\linewidth]{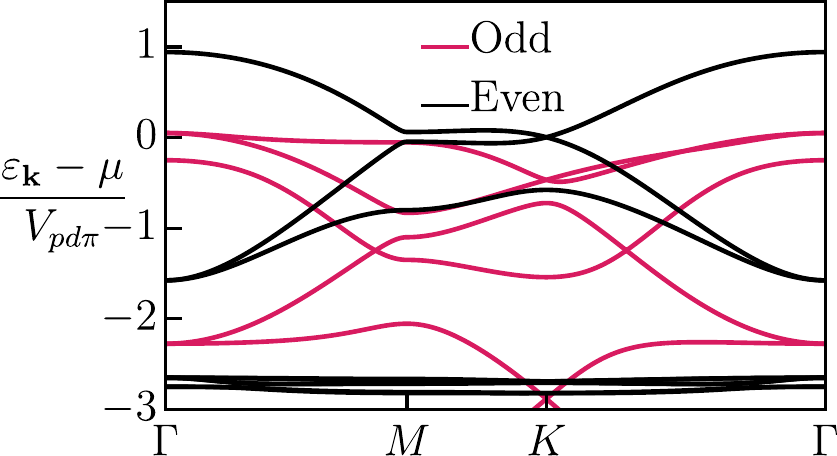}
    \caption{
    Interference-induced nearly flat  bands remain in a multi-orbital kagome-Lieb lattice  model, with $d$-orbitals on the metals and $\pi$-orbitals on the ligands \cite{sup}. Here we show the model for parameters chosen to reproduce the key features of  Cu-BHT (cf. Ref.~\cite{Cu-BHT-theo}). $V_{pd\pi}\sim1${~eV} is a Slater-Koster energy integral \cite{sup,SlaterKoster,koskinen2009}.}
    \label{fig:multiorb}
\end{figure}

\begin{figure}
    \centering
    \includegraphics[width=0.9\linewidth]{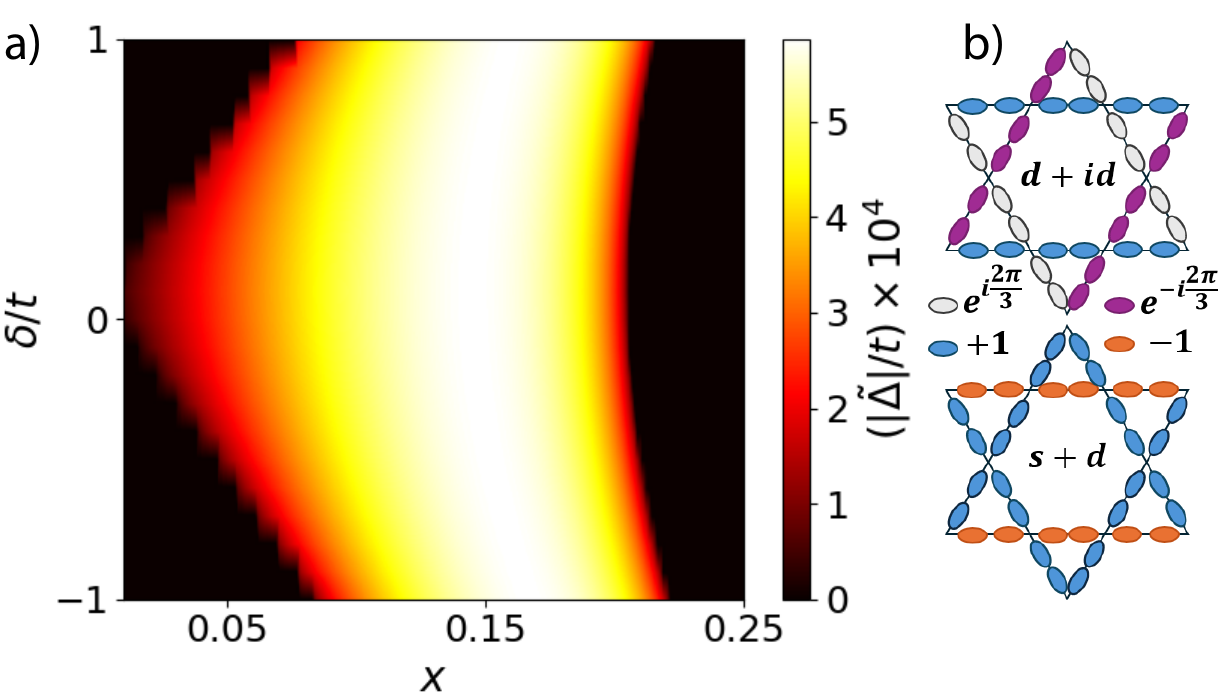}
    \caption{
        (a) Three degenerate  superconducting phases emerge on doping from  half-filling ($s$, $s+d$ and $d+id$).  Here $J=0.2t$, the inverse temperature is $\beta t = 2000$, $\tilde{\Delta}$ is the superconducting order parameter, and $x$ is the hole doping per unit cell. (b) Sketches of the non-trivial superconducting order parameters.
        }
    \label{fig:supercond}
\end{figure}

In this Letter, we argue that reticular materials generically form bipartite lattices with flat bands, which can be understood as a consequence of Sutherland's theorem \cite{SutherlandBill1986Loew,Calugara2021}. We explore a simple example; the kagome-Lieb lattice,  \cref{fig:kagomelieb}a, which is to the kagome lattice as the Lieb lattice is to the square lattice. We show that an $s$-orbital model on this lattice has five interference-induced flat bands (IFBs), three of which are partially occupied at half-filling, with large gaps between them and all other bands \cref{fig:kagomelieb}b. We show that, and explain why, they are robust to the largest expected perturbations, Figs. \ref{fig:KLbalents}, \ref{fig:KLBS}. The multi-orbital analog has a closely related band structure with IFBs, Fig. \ref{fig:multiorb}, which can be understood in terms of the s-orbital model, and reproduces the key features of the band structure of the superconducting MOF,  Cu-BHT, in which the metals form a kagome lattice with ligand sites along the bonds. When strong electronic correlations are included, we find that the $s$-orbital model leads to three degenerate  superconducting phases, Fig. \ref{fig:supercond}.

CPs consist of metals bonded to  organic groups (ligands) \cite{YaghiSciAdv}. These ligands link to other metals to form a crystal. 
The large unit cells of CPs mean that beyond nearest-neighbor hopping is often negligible, resulting in bipartite lattices.

For a $d$-dimensional CP with periodic boundary conditions, $|N_L-N_M|\geq d-1$, where $N_L$ ($N_M$) is the number of ligand (metal) sites per unit cell  \cite{sup}. Whence Sutherland's theorem \cite{SutherlandBill1986Loew}  requires that the nearest-neighbor tight-binding model for a reticular material must contain at least $d-1$ degenerate flat bands. This makes reticular materials playgrounds for flat bands.

MOFs are the subclass of CPs with high porosity  \cite{YaghiSci}, and hence have widespread potential applications, such as carbon capture and storage, hydrogen fuel storage, microelectronics, and drug delivery \cite{YaghiSci,Dinca-review,Deanna-review}. 
COFs are similar to CPs and MOFs, but with the metals replaced by an organic group \cite{COFreview,COF-review}. Therefore, one also expects zero energy flat bands in MOFs and COFs.
`Crystal  engineering' \cite{Robson1,Robson2}: the rational design of the crystal structures of reticular materials is enabled  by the local nature of chemical bonding \cite{YaghiSciAdv,YaghiNature,Robson1,Robson2,COF-Lieb}. Thus, realizing specific lattice structures in reticular materials is a highly achievable objective.
Reticular lattices could also be realized in optical lattices.

\textit{The kagome-Lieb lattice.-}To substantiate these ideas, we  explore an example. In many CPs and MOFs \cite{CuBHT-ACIE,TakenakaT2021Scsi,PeterJacobsonstructure,kgmTBAMOFQiMeiling2023DCo2,KLlargecentralligandChenZhijun2020Noan,Hau-kagomeLieb,Dong-kagomeLieb,Dinca-kagome} the metals form a kagome lattice with the ligands lying along the nearest-neighbor bonds. For example, density functional theory (DFT) calculations show that the low-energy electronic structure of Cu-BHT and  is dominated by bands that originate from the Cu and S atoms with little weight on the C atoms \cite{Cu-BHT-theo}; structurally, the S atoms sit along the nearest-neighbor Cu-Cu bonds. As both the metals and ligands have  orbitals near the Fermi energy and the s-orbital tight-binding approximation is applicable to many MOFs \cite{SilveiraOrlandoJ2016ESo2, FieldBernard2022Cmis, KumarAvijit2018TBSi, NiXiaojuan2020πYKb},  the simplest possible model is the tight-binding model on the kagome-Lieb lattice, \cref{fig:kagomelieb}a. Importantly, increasing covalency between metals and ligands is a  design strategy for achieving metallic and superconducting MOFs \cite{Dinca-review}, which motivated the first studies of Cu-BHT \cite{TakenakaT2021Scsi}, and necessitates electronic models including both the metals and the ligands. This lattice also occurs quite naturally in COFs \cite{COFreview}.
In contrast, the kagome-Lieb lattice is absent from recent catalogs of flat-band inorganic materials \cite{BernevigCatalogue, NevesPaulM.2024Cnco, LiuHang2021Stmw}.
Thus, we study the Hamiltonian 
\begin{equation}
    H_{tb}= -\sum_{i,j,\mu,\nu, \sigma} t_{ij}^{\mu\nu} \hat{c}_{i\mu\sigma}^{\dagger}\hat{c}_{j\nu\sigma}   +  \sum_{i\in M,\mu,\sigma} \delta_\mu \hat{c}_{i\mu\sigma}^{\dagger}\hat{c}_{i\mu\sigma},
    \label{eq:H}
\end{equation}
where $\hat{c}_{i\mu\sigma}^{(\dagger)}$ creates (annihilates) an electron of spin $\sigma$ in orbital $\mu$ at site $i$, $t_{ij}^{\mu\nu}$ is the hopping amplitude between orbitals on different sites, and the sum over $i\in M$ includes metal sites only. 

Generically, reticular materials will be multi-orbital, but we can learn a lot about the kagome-Lieb lattice  by considering the simplest model with one $s$-orbital per site: we suppress the orbital indices and sketch the hopping amplitudes  $t_{ij}^{\mu\nu}\in\{t,t_M,t_{LH},t_{LT}\}$ in \cref{fig:kagomelieb}a.  

In the nearest-neighbor model ($\delta=t_M=t_{LH}=t_{LT}=0$) there are  five IFBs:  three degenerate IFBs at $\varepsilon_{\bm{k}}=0$ and two `kagome-like' IFBs, which touch  kagome-like  dispersive bands (\cref{fig:kagomelieb}b). 
The three degenerate flat bands are required by 
Sutherland's theorem \cite{SutherlandBill1986Loew} for $N_L=6$, $N_M=3$; however, the kagome-like flat bands are not.

\textit{Localised states.-}The five flat bands result from quantum interference due to the topology of the lattice and are thus IFBs  \cite{BernevigReview,BergmanDoronL2008Btfr,Hwang,Jacko-flat,Mallah}. 
The localized states that comprise the flat-band of the kagome lattice consist of alternating sign amplitudes around each hexagon \cite{BergmanDoronL2008Btfr}. The  kagome-like IFBs on the kagome-Lieb lattice  are composed of the (anti)bonding combination of the metals and two nearest-neighbor ligands on hexagon $h$. These  hybrid orbitals  form  flat bands,  ($|+\rangle_h$) $|-\rangle_h$, analogous to those found in the kagome lattice (\cref{fig:KLbalents}a,b). The dispersive bands with (high) low energies are also  kagome bands in the basis of the (anti)bonding states of metals/nearest-neighbor ligands. This explains the Dirac point at K (\cref{fig:kagomelieb}b). In contrast, the three degenerate IFBs are non-bonding (between metals and ligands),
as they have no weight on the metal sites (\cref{fig:KLbalents}c-e). The non-bonding states can also be understood as superpositions of `bow-tie' states \cref{fig:KLbalents}f. 

The two kagome-like flat bands are singular as they touch a dispersive band at the $\Gamma$ point ($k=0$). This has a simple explanation, analogous to that for the kagome \cite{BergmanDoronL2008Btfr} and other decorated kagome lattices \cite{Jacko-flat}. Each unit cell contains one hexagon (\cref{fig:kagomelieb}). 
Therefore, for  $N$ unit cells and open boundary conditions (OBC) there are $N$ (anti)bonding kagome-like IFB states. Superpositions of $|\pm\rangle_h$ states on adjacent hexagons have zero weight on the sites that contribute to both localized wavefunctions due to destructive interference. 
Therefore,  $\sum_h|+\rangle_h=\sum_h|-\rangle_h=0$ for periodic boundary conditions (PBC), leaving $N-1$ non-trivial superpositions of  $|+\rangle_h$ ($|-\rangle_h$).
However, for PBC one can construct two additional states, which cannot be written as superpositions of $|+\rangle_h$ ($|-\rangle_h$), and wrap around the lattice (i.e., one noncontractable loop composed of (anti)bonding states in each of the $x$ and $y$ directions) \cite{BergmanDoronL2008Btfr,Jacko-flat}; resulting in $N+1$ states with energy $E_\pm$. As each IFB can only accommodate $N$  states, topology dictates that there cannot be a gap between the  kagome-like IFBs and the dispersive bands.

The non-bonding states are most easily counted in the basis of bow-tie states  (\cref{fig:KLbalents}f). There are three metal sites, and thus three linearly independent bow-tie states, per unit cell. All noncontractable loops are superpositions of bow-ties (\cref{fig:KLbalents}g). Thus, the $3N$ bow-tie states are non-singular and need not touch a dispersive band. 

The non-singularity of the non-bonding IFBs allows significant gaps between the non-bonding IFBs and the dispersive bands. In contrast, the IFBs on the kagome \cite{BergmanDoronL2008Btfr}, Lieb \cite{Hwang}, pyrochlore \cite{BergmanDoronL2008Btfr}, and star \cite{Jacko-flat} lattices are all singular and have topologically required band touchings, and therefore, excitations into dispersive bands occur at arbitrarily low energies. Thus, materials realizing these lattices are significantly less promising candidates to explore flat band physics at high electronic densities than materials that form kagome-Lieb lattices. New reticular materials tailored to exploit this structure could  be readily synthesized to complement  existing examples, such as Cu-BHT.  

\textit{Beyond nearest-neighbor hopping.-} A mismatch in the orbital energies of the metals and ligands, $\delta\ne0$, does not induce a curvature in the degenerate IFBs as the lattice remains bipartite, (and, more intuitively, none of the localized wavefunctions contain any amplitude on the metals, Fig. \ref{fig:KLbalents}c-e) \cite{SutherlandBill1986Loew}.  The two kagome-like IFBs remain, as $\delta$ simply changes the amplitudes on the metal and ligand orbitals in the (anti)bonding orbitals.
However, $\delta\ne0$ breaks particle-hole symmetry;  positive $\delta$  increases (decreases) the energy gap between the non-bonding IFBs and the (anti)bonding kagome-like bands, $E_g^{(h)}$ ($E_g^{(e)}$){, \cref{fig:KLBS}a}.

Direct metal-metal hopping, $t_{\text{M}}$, does not cause any of the five IFBs of the kagome-Lieb lattice to become dispersive, \cref{fig:KLBS}b. This is surprising as the lattice is no longer bipartite for  $t_M\ne0$. Therefore, Sutherland's theorem no longer guarantees that the model retains the three non-bonding IFBs.  However, as the non-bonding IFBs have zero amplitude on all metal sites, metal-metal hopping cannot perturb them. The two kagome-like IFBs have non-zero amplitudes on the metal sites. However, in  $|+\rangle_h$  and $|-\rangle_h$, the phases of the metal sites alternate around each hexagon  and are therefore localized by destructive interference under direct metal-metal hopping, \cref{fig:KLbalents}. Therefore, the contribution to the energy of kagome-like IFBs from metal-metal hopping is independent of momentum and these bands remain completely flat.

Thus, the five IFBs remain completely dispersionless, regardless of the values of $\delta$ and  $t_{\text{M}}$, \cref{fig:KLBS}a-c. This is important because the values for $\delta$ and $t_{\text{M}}$ in CPs and MOFs can vary dramatically, even between closely related materials \cite{KennyE.P2021TAtP}. 

Direct ligand-ligand hopping is typically small; however, it is not always negligible \cite{Fuchs_CRPA,Bernard}. We consider ligand-ligand hoppings across triangles, $t_{LT}$, and  across hexagons, $t_{LH}$; \cref{fig:kagomelieb}a. For example, the latter could be mediated by virtual hoping via the S-$p_z$ orbitals in Cu-BHT. Both have important qualitative effects on the band structure of the kagome-Lieb lattice, \cref{fig:KLBS}d-f. If exactly one of $t_{LT}$ and $t_{LH}$ is non-zero only one of the non-bonding IFBs remains exactly flat. This can be understood  from interference and lattice topology as a simple extension of the above arguments \cite{sup}. 

For both $t_{LT}\ne0$ and $t_{LH}\ne0$, no completely flat bands remain, \cref{fig:KLBS}f. However, the non-bonding bands remain nearly flat and widely gapped, indeed, significantly flatter and more isolated than the nearly flat bands in TBG, \cref{fig:KLBS}\spctwo g,h. This model reproduces the key features of the electronic structure of the MOF Cu$_3$C$_6$O$_6$ \cite{Shaiek}, which is chemically analogous to Cu-BHT with O substituted for S.

The large unit cells of reticular materials mean that longer range hopping is likely to be  negligible. Thus we expect that these conclusions will hold in many framework materials  described by kagome-Lieb lattice tight-binding models.

It is also interesting to anticipate future experiments where a 2D kagome-Lieb material is grown on a substrate. Strain caused by the substrate will perturb the ideal lattice and hence the hopping paramaters. The non-bonding IFBs will  be robust to such distortions because they do not change the bipartite nature of the lattice so Sutherland's theorem still holds.

\textit{Multi-orbital tight-binding model- }The multi-orbital tight-binding model (Eq. \ref{eq:H}) with a Slater-Koster parameterization \cite{sup} reproduces
the key features of the DFT band structure of Cu-BHT \cite{Cu-BHT-theo}, \cref{fig:multiorb}.
The multi-orbital tight-binding model hosts several flat and nearly flat bands, which can readily be understood from simple generalizations of the arguments above. The model is symmetric under reflection through the plane and hence is separable into even (odd) sectors, with support from ligand $\pi_x$ and $\pi_y$ orbitals, and metal $d_{xy}$, $d_{x^2-y^2}$ and $d_{z^2}$ (ligand $\pi_z$ and metal $d_{xz}$ and $d_{yz}$) orbitals.
These sectors are  decoupled at the single-electron level, but there are Coulomb interactions between electrons in the different sectors.
The $d_{z^2}$ orbitals are energetically separated and do not hybridize significantly with the other orbitals in Cu-BHT; therefore we neglect them below. This leaves 12 (6) ligand and 6 (6) metal orbitals per unit cell in the even (odd) sector. Therefore, we expect 6 even non-bonding IFBs with support on the $\pi_x$ and $\pi_y$ orbitals (\cref{fig:multiorb}).  

There are twice as many IFBs in the even sector of the multi-orbital model as there are in the $s$-orbital model because there are two orbitals per site in the even sector. 
The compact localised states that give rise to the IFBs in the multi-orbital model are simply related to those that give rise to the non-bonding IFBs in the s-orbital model: the single orbitals are replaced by a superposition of the two orbitals on a given site.
This leads to a deep connection between the resulting IFBs in the two models: both are required by Sutherland’s theorem, are robust to strain, remain exactly flat when metal-metal hopping is added even though the lattice is no longer bipartite, and remain very narrow when ligand-ligand hopping is added \cite{sup}.

\textit{Superconductivity.-}
One expects metals with partially occupied flat bands to be unstable. Various symmetry breaking or topological orders can occur, including the fractional quantum Hall effect, magnetic states, charge density waves, or superconductivity. Determining, theoretically, which is realized is extremely challenging. Nevertheless, to illustrate the role of electronic correlations, we study the nearest-neighbor $s$-orbital kagome-Lieb $t$-$J$ model:

\begin{equation}
    {H_{tJ}= \hat{P} \Big[
        H_{tb} + J \sum_{\langle i,j\rangle_1} \Big( \hat{\bm{S}}_i \cdot \hat{\bm{S}}_j - \frac{1}{4} \hat{n}_i \hat{n}_j \Big)
        \Big] \hat{P},}
    \label{eq:tJ}
\end{equation} 

where $\hat{P}$ is the Gutzwiller projector and $\hat{\bm{S}}_i$ ($\hat{n}_i$) is the spin (number) operator on site $i$. We write the electronic operators $\hat{c}_{i\sigma}=\hat{b}_i^\dagger \hat{f}_{i\sigma}$ in terms of (slave) bosonic holons, $\hat{b}_i$, and fermionic quasiparticles, $\hat{f}_{i\sigma}$ \cite{KotliarRuckenstein} 
and calculate the superconducting order parameter, 
$\tilde{\Delta}_{ij}=\frac{J}{2}\langle \hat{c}_{i\uparrow} \hat{c}_{j\downarrow} - \hat{c}_{i\downarrow} \hat{c}_{j\uparrow} \rangle$, in the mean-field approximation \cite{KotliarLiu,sup}.

For non-zero $\delta$,  a charge transfer insulator \cite{Zaanen}  arises near half filling, due to the energy benefit of exclusively  doping holes onto either the ligands or the metals, depending on the sign of $\delta$. One sublattice remains exactly half-filled, thus charge transport is forbidden by the Gutzwiller projector in \cref{eq:tJ}.

Upon further hole doping we find three degenerate  superconducting states: $s$ ($A_1$), $s+d$  ($A_1+E_2$) and $d+id$ ($E_2$) characters, \cref{fig:supercond}b. High degeneracy is not unexpected in a flat band system, but it is interesting that it survives in the strongly correlated limit considered here.
The $d+id$  breaks time reversal symmetry. The $s$ state is fully gapped. The $d+id$ and $s+d$ states have line (surface) nodes in 2D (3D) \cite{Powell2007,Powell2006}
at optimal doping.  However, these nodes are formally accidental, and lift away from optimal doping. Interestingly, there is experimental evidence for nodes in Cu-BHT \cite{TakenakaT2021Scsi}. 
Non-zero $\delta$ suppresses superconductivity at  low doping, but not near optimal doping, suggesting that superconductivity in reticular materials can be robust to energy differences between the metals and ligands.

While $\tilde{\Delta} \neq 0$  indicates condensing Cooper pairs, a non-zero superfluid stiffness is required to exhibit the Meissnner effect.
For an isolated set of partially filled flat bands the superfluid stiffness in the weak coupling limit is given by $D_s \propto \text{tr}(g) |\tilde{\Delta}|$, where $g$ is the minimal Fubini-Study quantum metric \cite{Torma2015,Torma2016,Torma2017,Torma2018,BernevigReview,Torma2022}.
The equivalent condition for strong coupling superconductivity remains an open question. We find $\text{tr}(g) = 0.14$  for the parameters in \cref{fig:KLBS}f. Thus, we expect a non-zero superfluid stiffness in the superconducting phases.
For $t=1$~eV, which is comparable to Cu-BHT (Fig. \ref{fig:multiorb}), and $J=0.2t$, we find $|\tilde{\Delta}|=7$~K at optimal doping, somewhat larger than the critical temperature of Cu-BHT (0.25--0.3~K) \cite{TakenakaT2021Scsi}.

\textit{Conclusions.-}This demonstrates that the very narrow, isolated bands in materials with the kagome-Lieb lattice exhibit exotic, strongly correlated phenomena, including unconventional superconductivity.  Thus, the kagome-Lieb lattice is an unusually rich playground for unconventional superconductivity. This provides strong support for our central hypothesis that reticular materials are an ideal platform for investigating flat band physics. A key design principle, derived from the results above, is to choose ligands with low direct ligand-ligand overlap as this minimizes the bandwidth of the non-bonding bands.

This work was supported by the Australian Research Council (DP180101483) and MEXT Quantum Leap Flagship Program (MEXT Q-LEAP) Grant Number JPMXS0118069605.

%

\clearpage

\renewcommand\thefigure{S\arabic{figure}}
\setcounter{figure}{0}  
\renewcommand\thetable{S\arabic{table}}
\setcounter{table}{0}  
\renewcommand\theequation{S\arabic{equation}}
\setcounter{equation}{0}  
\end{document}


\title{Supplementary Information: Flat bands and unconventional superconductivity in a simple model of metal-organic frameworks}

\author{M. F. Ohlrich}
\email{miriamohlrich@gmail.com}
\affiliation{School of Mathematics and Physics, The University of Queensland, 4072, Australia}

\author{E. M. Makaresz}
\affiliation{School of Mathematics and Physics, The University of Queensland, 4072, Australia}

\author{H. L. Nourse}
\affiliation{Quantum Information Science and Technology Unit, Okinawa Institute of Science and Technology Graduate University, Onna-son, Okinawa 904-0495, Japan}
\thanks{Present Address: School of Chemistry, University of Sydney, NSW 2006, Australia}

\author{B. J. Powell}
\email{powell@physics.uq.edu.au}
\affiliation{School of Mathematics and Physics, The University of Queensland, 4072, Australia}

\maketitle

\renewcommand\thefigure{S\arabic{figure}}
\setcounter{figure}{0}  
\renewcommand\thetable{S\arabic{table}}
\setcounter{table}{0}  
\renewcommand\theequation{S\arabic{equation}}
\setcounter{equation}{0}  

\section{Consequences of Sutherland's Theorem for Liebified lattices}

\begin{figure}
    \centering
    \includegraphics[width=0.85\linewidth]{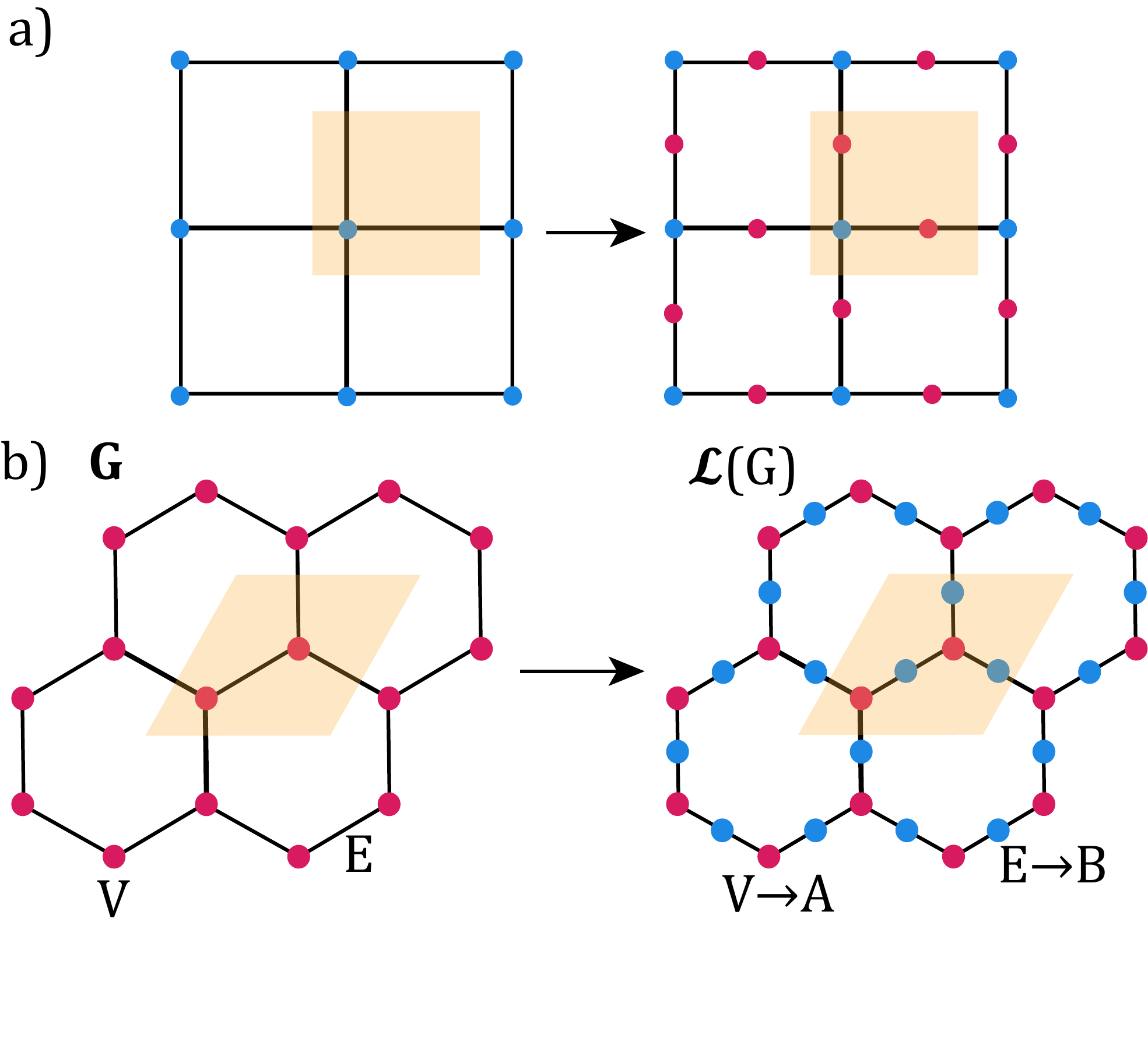}
    \caption{When an additional site is placed at the midpoint of each bond on any lattice, the resulting lattice is `Liebified', and is bipartite. a) The Lieb lattice arises from Liebifing the square lattice. b) Liebifing the honeycomb lattice results in the Lieb-honeycomb lattice (commonly known as honeycomb-kagome, which is realised by many reticular materials). The original lattice is the periodic tiling of a graph $G$ (unit cell in yellow), with vertices $V$ and edges $E$. The corresponding Liebified lattice, $\mathcal{L}(G)$, is the mapping of $V$ and $E$ in $G$ to the $A$ and $B$ sites. Clearly, $\mathcal{L}(G)$ is bipartite for nearest-neighbor hopping only.}
    \label{fig:liebification}
\end{figure}

The Lieb lattice can be constructed by starting from the square lattice and adding a site at the midpoint of every bond, Fig. \ref{fig:liebification}a.
Henceforth, we will refer to any lattice that can be produced by starting from another lattice and adding a site at the midpoint of every bond as a Liebified lattice, Fig. \ref{fig:liebification}.

Reticular materials are made up of two distinct building blocks: e.g., in MOFs, the ligands form sublattice $A$ and the metals form sublattice $B$. The large unit cells of reticular materials mean that beyond nearest-neighbor hopping is often negligible; therefore, reticular materials  commonly form bipartite lattices. More specifically, reticular materials generally form Liebified lattices, with the $A$-subunits placed along the bonds of the original lattice (the $B$-subunits). For  an s-orbital nearest-neighbor tight-binding model on a bipartite lattice, Sutherland's theorem \cite{SutherlandBill1986Loew} states that there must be $N_F\geq |N_A - N_B|$ degenerate flat bands, where $N_A$ is the number of sites in sublattice  $A$ and $N_B$ is the number of sites in sublattice $B$. 

We now show that $N_A \neq N_B$ for any Liebified lattice in more than one dimension. Thus, Sutherland's theorem implies that reticular materials in two or more dimensions will display flat bands. For the following argument, we define a lattice to be an undirected, connected graph $G(V, E)$ with periodic boundary conditions, $N_V$ vertices (sites), and $N_E$ edges (bonds) per unit cell \cite{WilsonRobin2010GraphTheoryIntro, DiestelReinhard2006GraphTheoryIntroTechnical}. Where edges (vertices) cross the unit cell boundary, they contribute fractionally to $N_E$ ($N_V$), always adding to an integer. We only  consider connected graphs, because the electronic structure of unconnected subunits (e.g., guest molecules) can be determined independently. We  allow $G$ to be a general graph, which may contain loops or multiple edges \cite{GraphTheoryGrossJonathanL2006Gtai}, and the loops and multiple edges can form cycles \footnote{The chemical richness of reticular materials allows for loops and multiple edges, e.g., via bidentate ligands or where two ligands connect a pair of metals, respectively.}.

Liebifing a graph $G$ maps the numbers of edges and vertices in $G$  to the numbers of vertices in each sublattice of the Liebified lattice, $\mathcal{L}(G)$, i.e., $N_A=N_E$, and $N_B=N_V$, \Cref{fig:liebification}. For a connected graph, $N_E=N_V$ if and only if the graph is unicyclic, i.e., contains exactly one cycle \cite{GraphTheoryGrossJonathanL2006Gtai, anderson1967unicyclic, pseudotreedefGabowHaroldN.1988Alaf}.  

The number of distinct cycles in a graph is $N_C\geq N_E-N_V+\xi$, where $\xi$ is the number of connected components of the graph \footnote{A connected component is a connected subgraph that is not part of any larger connected subgraph.}, see \Cref{sec:proof} for a proof \cite{GraphTheoryandApplicationsBondy_orange, GraphTheoryGrossJonathanL2006Gtai}. For any connected graph, $\xi=1$, so if $G$ contains two or more cycles, then $N_E>N_V$, leading to an associated Liebified lattice ${\cal L}(G_d)$ with $N_A>N_B$. 
 
A one-dimensional lattice must contain at least one cycle: wrapping around the perimeter. A two-dimensional lattice must have at least two cycles: one wrapping around each direction of the torus. As adding an additional periodic boundary  adds at least one cycle, a $d$-dimensional lattice with periodic boundary conditions, $G_d$, must contain at least $d$-cycles. As $\xi=1$ for any lattice,  $N_E-N_V\geq d-1$. Equivalently, the corresponding Liebified lattice, ${\cal L}(G_d)$, has $N_A-N_B\geq d-1$. Therefore, the nearest-neighbor tight-binding model for  ${\cal L}(G_d)$ must contain at least $d-1$ degenerate flat bands. 
Consequently, the nearest-neighbor tight-binding models for all s-orbital Liebified lattices in two or three dimensions must have at least one flat band.

One-dimensional lattices only have one cycle due to periodicity, and so, in general, Liebified one-dimensional lattices are not guaranteed to have flat bands. However, if one or more cycles are present within the unit cell of original lattice, e.g., the triangular necklace \cite{TriangularnecklaceJananiC.2014Leto} or Aharonov-Bohm caging in the diamond lattice \cite{Butaud2000Aharonov}, then at least one zero-energy flat band will be present in the Liebified lattice. 

The above trivially generalizes to non-periodic lattices, $G_f$. Whence one finds that if the Liebified lattice ${\cal L}(G_f)$ has at least $n-1$ degenerate states with the energy of the majority site if $G_f$ has $n$ cycles.

This makes reticular materials playgrounds for flat bands.

\subsection{Proof that \texorpdfstring{$N_C \geq N_E-N_V+\xi$}{}}
\label{sec:proof}

\textbf{Lemma: At least one cycle is created if a single edge is added to an undirected, connected graph $G(V,E)$, where multiple edges and loops are permitted.}

There are two possible ways to add an edge to $G$. 
Firstly, a loop, $e(u, u)$ which connects a vertex $u$ to itself, can be added. A loop is a cycle \cite{GraphTheoryGrossJonathanL2006Gtai}, therefore, adding a loop creates a cycle.
Secondly, an edge $e(u,v)$, which connects vertex $u$ to vertex $v$ can be added. As $G$ is connected, there exists a path between all pairs of vertices \cite{GraphTheoryandApplicationsBondy_orange}, so there is already a path between $u$ and $v$. Adding $e(u,v)$ creates an additional path between $u$ and $v$, thus forming a cycle. Therefore, if a single edge is added to $G$, then at least one cycle is created.

\textbf{Proof: An undirected graph $G$, where multiple edges and loops are permitted, contains at least $N_E-N_V+\xi$ cycles, where $N_E$ is the number of edges, $N_V$ is the number of vertices, and $\xi$ is the number of connected components of $G$}

A connected component, $i$, of $G$ has $N_{V_i}$ vertices. Therefore, the total number of vertices in $G$ is 

\begin{equation}
N_{V}=\sum_{i=1}^\xi N_{V_i}. 
\end{equation}

The spanning tree of $i$ has $N_{E_{T_i}}=N_{V_i}-1$ edges \cite{GraphTheoryandApplicationsBondy_orange}. So, the total number of edges in all of the spanning trees across $G$ is

\begin{align}
N_{E_T}  &=\sum_{i=1}^\xi  (N_{V_i} - 1) \\
&=\sum_{i=1}^\xi  N_{V_i} -\sum_{i=1}^\xi 1\\
&=N_V - \xi.
\end{align}

The remaining edges $G$, $N_{E_{R}}$, that are not part of the spanning trees, must then equal

\begin{align}
N_{E_R}&=N_E-N_{E_T}\\
&=N_E-N_V+\xi.
\end{align}

As shown in the Lemma, if a single edge is added to a connected graph, then at least one cycle is created. A spanning tree is a connected graph, so this statement applies. Therefore, each edge in $G$ that is not part of the spanning trees results in the formation of at least one cycle. Thus, the number of cycles, $N_C \geq N_{E_R}$, so

\begin{align}
N_C \geq N_E-N_V+\xi.
\end{align}

\section{Compact localized states with ligand-ligand hopping}

For $t_{LT}\ne0$, $t_{LH}=0$,  two of the non-bonding  bands develop a weak dispersion, (Fig. 3d), and the energy of the remaining exactly flat non-bonding band shifts to $E_{n_1}= t_{\text{LT}}$. The two kagome-like IFBs also remain flat. We can understand why these bands  remain flat by examining their localized wavefunctions, Fig. 2a-c.
First consider the model with $t_{LT}\ne0$ and all other parameters set to zero. The model now consists of disconnected triangles of ligand sites. A single triangle has a bonding orbital with energy $-2t_{LT}$, and two antibonding orbitals of energy $t_{LT}$ \cite{NaCoO2}. $|n_1\rangle_h$ is a linear superposition of these triangular antibonding orbitals on six triangles; therefore it is an eigenstate with energy $t_{LT}$. 
We have seen above that $|n_1\rangle_h$ is a zero energy eigenstate of the terms proportional to $t$, $\delta$, and $t_m$.
Therefore, $|n_1\rangle_h$ is an eigenstate of the full model with $t_{LH}=0$ with energy $t_{LT}$.
A similar argument explains why $|+\rangle_h$ and $|-\rangle_h$, which can be viewed as superpostions of the triangular antibonding states and atomic metal states, are eigenstates.
$|n_2\rangle_h$ and $|n_3\rangle_h$ are no longer eigenstates, as ligand-ligand hopping across a triangle results in a non-zero amplitude on  sites where the wavefunctions previously vanished,  Fig. 2d,e.

For $t_{LT}=0$, $t_{LH}\ne0$, only one of the non-bonding bands remains exactly flat, with an energy of $E_{n_3}=2 t_{\text{LH}}$, Fig. 3e.
This can be readily understood, as the model with $t_{LH}\ne0$ and all other parameters set to zero consists of disconnected hexagons connecting ligand sites; and $|n_3\rangle_h$ is highest energy antibonding state of the tight-binding model on a single hexagon, Fig. 2e \cite{Lowe}.
The other four localized states are not eigenstates for $t_{LH}\ne0$, Fig. 2a-d.

\section{Slave boson theory of superconductivity}

The slave boson theory detailed below extends the procedure outlined in \cite{KotliarRuckenstein, KotliarLiu} for the resonating valence bond (RVB) theory on the $t$-$J$ model, Eq. (2), with isotropic nearest-neighbor hopping and one $s$-orbital per site. Here we generalize this to treat inequivalent sites within the unit cell, as necessary for the kagome-Lieb lattice. 

A spinless auxiliary boson, $\hat{b}_{i}$, and fermion of spin $\sigma$, $\hat{f}_{i \sigma}$, are introduced for each site $i$. A faithful representation of the electronic operators in the subspace with no doubly occupied sites is $\hat{c}_{i \sigma} = \hat{b}_i^{\dagger} \hat{f}_{i \sigma}$. With the interpretation that the bosons occupy only sites empty of electrons, the Gutzwiller projector is represented as a set of constraints,
\begin{equation}
    \sum_{\sigma} \hat{f}^{\dagger}_{i \sigma} \hat{f}_{i \sigma} + \hat{b}^{\dagger}_i \hat{b}_i = 1,
\end{equation}
which are enforced by Lagrange multipliers, $\lambda_i$, on each lattice site. As in Ref. \cite{KotliarLiu} the Lagrange multipliers are averaged over the unit cells to make the problem tractable, so that only nine independent Lagrange multipliers are used, one for each site within the unit cell.

A Hartree-Fock-Gorkov mean-field factorisation is performed, introducing the RVB, $\Delta_{ij\sigma}=\langle \hat{f}_{i\sigma} \hat{f}_{j\Bar{\sigma}} \rangle$, and quasiparticle nematic, $\chi_{ij\sigma}=\langle \hat{f}_{i\sigma}^\dagger \hat{f}_{j\sigma}\rangle$, order parameters, where $\Bar{\sigma}$ is the opposite spin to $\sigma$.  The RVB order parameter measures singlet pairing, so we enforce $\Delta_{ij} \equiv \Delta_{ij\uparrow} = -\Delta_{ij\downarrow}$. We only look for paramagnetic states, so we also enforce $\chi_{ij} \equiv \chi_{ij\uparrow} = \chi_{ij\downarrow}$. The density appears in the decoupling via $n_{i\sigma} = \langle \hat{f}^{\dagger}_{i\sigma} \hat{f}_{i\sigma} \rangle$, which can be combined into one parameter per site in the paramagnetic case, $n_i = \sum_{\sigma} n_{i\sigma}$. The bosons are assumed to  condense so that the boson operators are replaced with their expectation values.  Generically, the bosons will then be complex numbers, however they can be made real via the gauge transformation

\begin{equation}
    \hat{b}_{i} \rightarrow \hat{b}_{i} e^{i\phi_i}, \hspace{0.3in} \hat{f}_{i\sigma} \rightarrow \hat{f}_{i\sigma} e^{i\phi_i},
\end{equation}
which leaves the physical electron operators invariant. Finally, any terms of the form $1 + \hat{b}^{\dagger}_i \hat{b}_i$ are set to 1 as these terms are small near half filling, which is the region relevant to the RVB theory of superconductivity.

The resulting mean-field Hamiltonian is quadratic and can always be diagonalized. Doing so analytically on the kagome-Lieb lattice is impractical due to the dimension of the Hamiltonian, so the Hamiltonian is instead diagonalized numerically.  Expectation values of the auxiliary fermion operators, such as those that appear in the order parameters, are calculated from the elements of the unitary matrix that diagonalizes the Hamiltonian.  This allows for a simple self-consistent procedure following from an initial guess of $\Delta_{ij}$, $\chi_{ij}$ and the fillings $n_i$ while the total hole doping is fixed.

\subsection{Mean-field Hamiltonian, self-consistent equations and order parameter}

We assume that all unit cells are equivalent (homogeneity). Thus, the mean-field Hamiltonian in real-space is

\begin{multline}\label{eq:H_mf_realspace}
    H_{mf} = \sum_{\langle i\alpha, j\beta \rangle} \bigg[ -\bigg( t b_{\alpha} b_{\beta} + \frac{J}{2}\chi^*_{\alpha\beta} \bigg) \sum_{\sigma} \hat{f}^{\dagger}_{i\alpha\sigma} \hat{f}_{j\beta \sigma} \\
    - J\Delta_{\alpha\beta} \bigg( \hat{f}^{\dagger}_{i\alpha\uparrow} \hat{f}^{\dagger}_{j\beta\downarrow} - \hat{f}^{\dagger}_{i\alpha\downarrow} \hat{f}^{\dagger}_{j\beta\uparrow} \bigg) + h.c. \bigg] \\
    - \frac{J}{4}\sum_{\langle i\alpha, j\beta \rangle \sigma} \bigg( n_{\alpha} \hat{f}^{\dagger}_{j\beta\sigma} \hat{f}_{j\beta\sigma} + n_{\beta} \hat{f}^{\dagger}_{i\alpha\sigma} \hat{f}_{i\alpha\sigma} \bigg) \\
    - \sum_{i\gamma\sigma} \big( \mu + \lambda_{\gamma} \big) \hat{f}^{\dagger}_{i\gamma\sigma} \hat{f}_{i\gamma\sigma} + \delta \sum_{i\sigma, \gamma \in M} \hat{f}^{\dagger}_{i\gamma \sigma} \hat{f}_{i\gamma\sigma} \\
    + J\sum_{\langle i\alpha, j\beta \rangle} \bigg( \frac{1}{4} n_{\alpha} n_{\beta} + 2|\Delta_{\alpha\beta}|^2 + |\chi_{\alpha\beta}|^2 \bigg) \\
    - \sum_{i\gamma} \lambda_{\gamma} \big( b_{\gamma}^2 - 1 \big),
\end{multline}
where Latin subscripts label unit cells, Greek subscripts label sites within the unit cell, and $\mu$ is the chemical potential. 

After Fourier transforming, the self-consistent equations are

\begin{subequations}
    \begin{multline}
        \Delta_{\alpha\beta} = \frac{1}{2N} \sum_{\bm{k}} \bigg[ \langle \hat{f}_{-\bm{k}\beta\downarrow} \hat{f}_{\bm{k}\alpha\uparrow} \rangle e^{-i\bm{k}\cdot \bm{r}_{\alpha\beta}} \\
        - \langle \hat{f}_{\bm{k}\beta\uparrow} \hat{f}_{-\bm{k}\alpha\downarrow} \rangle e^{i\bm{k}\cdot \bm{r}_{\alpha\beta}} \bigg],
    \end{multline}
        
    \begin{equation}
        \chi_{\alpha\beta} = \frac{1}{2N} \sum_{\bm{k}\sigma} \langle \hat{f}^{\dagger}_{\bm{k}\alpha\sigma} \hat{f}_{\bm{k}\beta\sigma} \rangle e^{i\bm{k} \cdot \bm{r}_{\alpha\beta}},
    \end{equation}

    \begin{equation}
        n_{\gamma} = \frac{1}{N}\sum_{\bm{k}\sigma} \langle \hat{f}^{\dagger}_{\bm{k}\gamma\sigma} \hat{f}_{\bm{k}\gamma\sigma} \rangle,
    \end{equation}

    \begin{equation}
        b_{\gamma} = \sqrt{1 - n_{\gamma}},
    \end{equation}

    \begin{equation}
        \lambda_{\gamma} = -\frac{2t}{b_{\gamma}}\sum_{\langle \alpha, \beta \rangle} |\chi_{\alpha\beta}| \Big( \delta_{\gamma \alpha} b_{\beta} + \delta_{\gamma \beta} b_{\alpha} \Big) \cos{ \big( \phi^{\chi}_{\alpha\beta} \big) },
    \end{equation}
\end{subequations}
where $\phi^{\chi}_{\alpha\beta}$ is the phase of $\chi_{\alpha\beta}$, $N$ is the number of k-points for the Brillouin zone sampling, $\bm{r}_{\alpha\beta}$ is the real-space vector connecting sites $\alpha$ and $\beta$, $\delta_{\alpha\beta}$ is the Kronecker-delta, and $\langle \alpha, \beta \rangle$ denotes the sum over pairs of nearest neighbor sites.

The superconducting order parameter can be identified by performing the same mean-field decoupling on the physical electron operators, which gives 

\begin{equation}\label{eq:SC_order_param_definition}
    \begin{split}
        \Tilde{\Delta}_{\alpha\beta} &= \frac{J}{2N} \sum_{\bm{k}} \bigg[ \langle \hat{c}_{-\bm{k}\beta\downarrow} \hat{c}_{\bm{k}\alpha\uparrow} \rangle e^{-i\bm{k}\cdot \bm{r}_{\alpha\beta}} \\
        & \hspace{1in} - \langle \hat{c}_{\bm{k}\beta\uparrow} \hat{c}_{-\bm{k}\alpha\downarrow} \rangle e^{i\bm{k}\cdot \bm{r}_{\alpha\beta}} \bigg] \\
        &= Jb_{\alpha}b_{\beta} \Delta_{\alpha\beta}.
    \end{split}
\end{equation}

\subsection{Symmetry of order parameters on the kagome-Lieb lattice}
\label{sec:superconducting-symmetrry}

\begin{figure}
    \centering
    \includegraphics[width=0.6\linewidth]{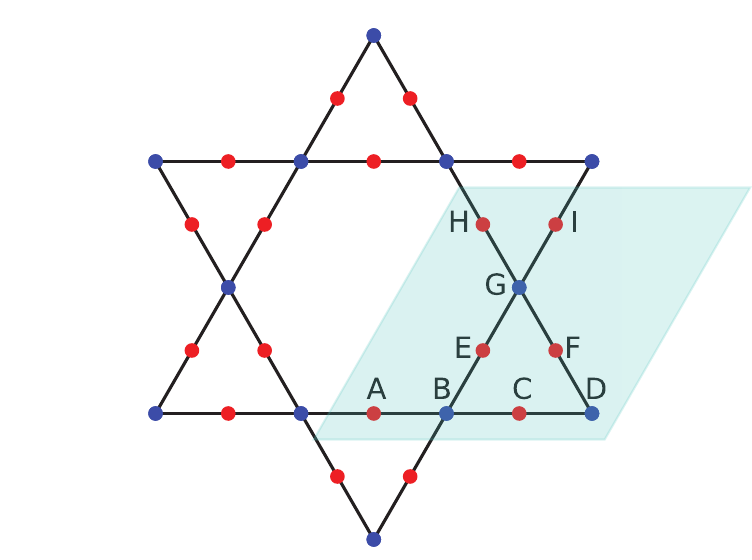}
    \caption{
    Site labels of the kagome-Lieb lattice used to define the basis symmetry vectors (cf. \cref{table:LK_irrep_symmetries}). The unit cell is shaded.
    }
    \label{fig:unitcell}
\end{figure}

Because the unit cell of the kagome-Lieb lattice contains many sites, it is useful to enforce symmetries of the order parameters to reduce the number of self-consistent parameters. This is represented mathematically as $\Delta_{ij} = \eta^{\Delta}_{ij} \Delta$ and $\chi_{ij} = \eta^{\chi}_{ij} \chi$, where the $\eta$ are the elements of the normalized `symmetry vector', which are fixed to constrain the symmetry of the order parameter. We assume all unit cells are equivalent and label the sites within each cell as shown in \cref{fig:unitcell}. A basis for the symmetry vectors is derived from the irreducible representations of the point group $C_{6v}$, which encompasses the symmetries of the kagome-Lieb lattice. By convention, these basis symmetries are labelled in analogy to the spherical harmonics ($s$, $p$, $d$, $\hdots$) and are shown in \cref{table:LK_irrep_symmetries}. The primed symmetries are a second orthogonal set of basis symmetry vectors corresponding to the same irreducible representation.
Arbitrary linear combinations of these basis vectors can also be considered, and mixing of different symmetries is allowed via, e.g. $\Delta_{ij} = \eta^{\Delta (1)}_{ij} \Delta_1 + \eta^{\Delta (2)}_{ij} \Delta_2$. 

\begin{table*}
\centering
\begin{tabular}{l | cccccccccccc}
    & $p_{x}$ & $p_{y}$ & $p'_{y}$ & $p'_{x}$ & $d_{xy}$ & $d_{x^2-y^2}$ & $d'_{x^2-y^2}$ & $d'_{xy}$ & $s$ & $i$ & $f_{x(x^2-3y^2)}$ & $f_{y(3x^2-y^2)}$ \\ \hline
    $\eta_{AB}$ & -1 &  1 & -1 &  1 & -1 &  1 & -1 & -1 &  1 &  1 &  1 &  1 \\
    $\eta_{BC}$ &  1 & -1 &  1 & -1 & -1 &  1 & -1 & -1 &  1 &  1 & -1 & -1 \\
    $\eta_{CD}$ & -1 & -1 &  1 &  1 &  1 &  1 & -1 &  1 &  1 & -1 &  1 & -1 \\
    $\eta_{DA}$ &  1 &  1 & -1 & -1 &  1 &  1 & -1 &  1 &  1 & -1 & -1 &  1 \\
    $\eta_{BE}$ &  2 &  0 & -2 &  0 & -2 &  0 &  2 &  0 &  1 & -1 &  1 & -1 \\
    $\eta_{EG}$ &  1 &  1 &  1 &  1 & -1 & -1 & -1 &  1 &  1 &  1 & -1 & -1 \\
    $\eta_{GI}$ & -1 & -1 & -1 & -1 & -1 & -1 & -1 &  1 &  1 &  1 &  1 &  1 \\
    $\eta_{IB}$ & -2 &  0 &  2 &  0 & -2 &  0 &  2 &  0 &  1 & -1 & -1 &  1 \\
    $\eta_{DF}$ & -2 &  0 & -2 &  0 &  2 &  0 &  2 &  0 &  1 &  1 & -1 & -1 \\
    $\eta_{FG}$ & -1 &  1 &  1 & -1 &  1 & -1 & -1 & -1 &  1 & -1 &  1 & -1 \\
    $\eta_{GH}$ &  1 & -1 & -1 &  1 &  1 & -1 & -1 & -1 &  1 & -1 & -1 &  1 \\
    $\eta_{HD}$ &  2 &  0 &  2 &  0 &  2 &  0 &  2 &  0 &  1 &  1 &  1 &  1 
\end{tabular}
\caption{
Unnormalised basis symmetry vectors of the kagome-Lieb lattice belonging to two-dimensional representations of the point group $C_{6v}$. 
Site labels within a unit cell are shown in \cref{fig:unitcell}.
}
\label{table:LK_irrep_symmetries}
\end{table*}

While we allow for nematic order in our calculations we always find that the ground states are isotropic ($s$ representation) where $\eta_{ij}^{\chi} = \frac{1}{\sqrt{12}} \,\, \forall \,\, i,j$. Hence, in the main text, we only report the superconducting order parameters in the phase diagram.

\subsection{Momentum dependence of the superconducting gap}

\begin{figure}
    \centering
    \includegraphics[width=\linewidth]{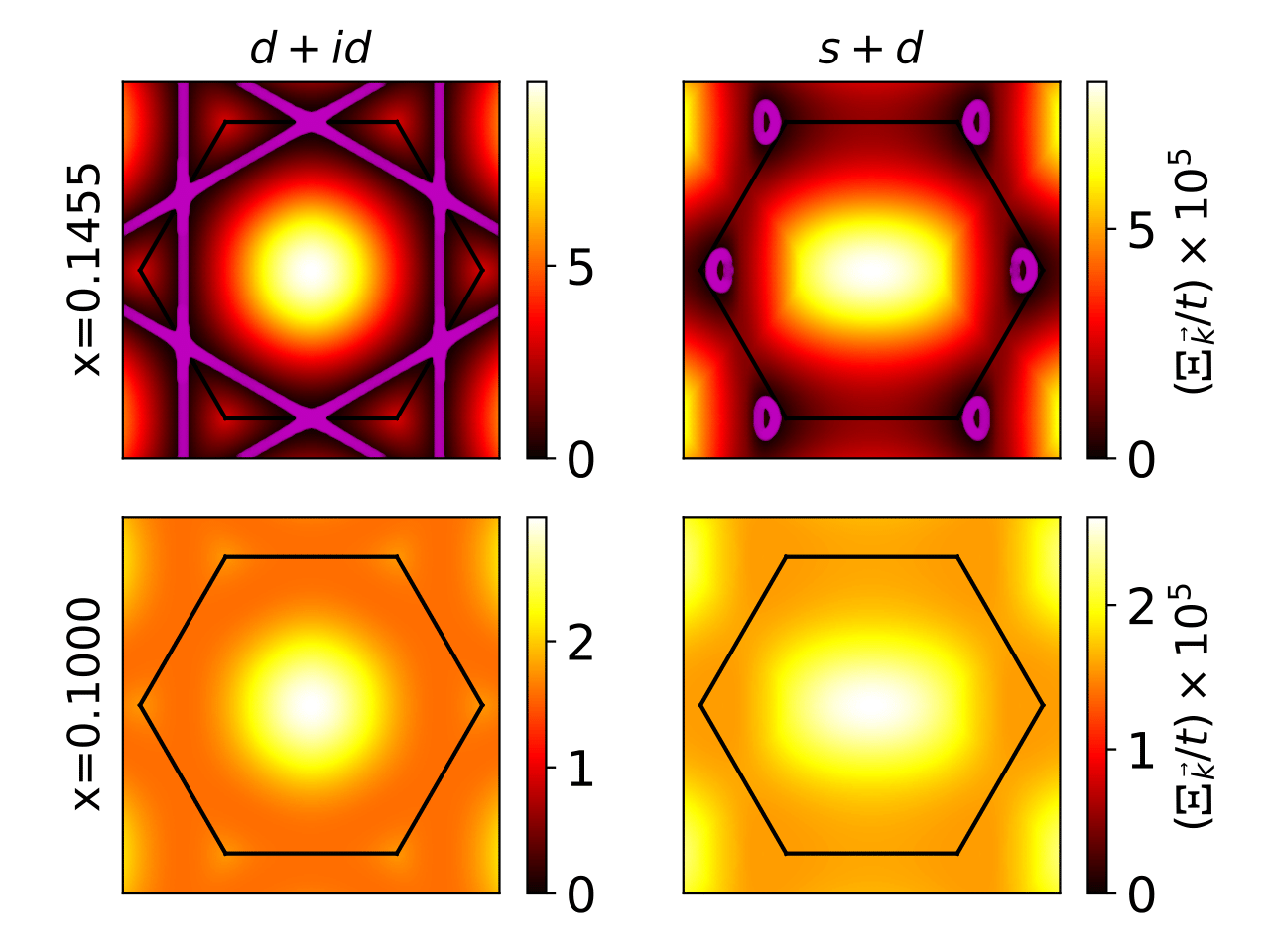}
    \caption{Momentum dependence of the superconducting gap, $\Xi_{\bm{k}}$, in the non-trivial symmetry states with $\delta = 0$ at optimal doping $x=0.1455$ per unit cell and away from optimal doping. The first Brillouin zone is outlined and the nodes are highlighted in magenta.}
    \label{fig:gap_k}
\end{figure}

\begin{figure}
    \centering
    \includegraphics[width=\linewidth]{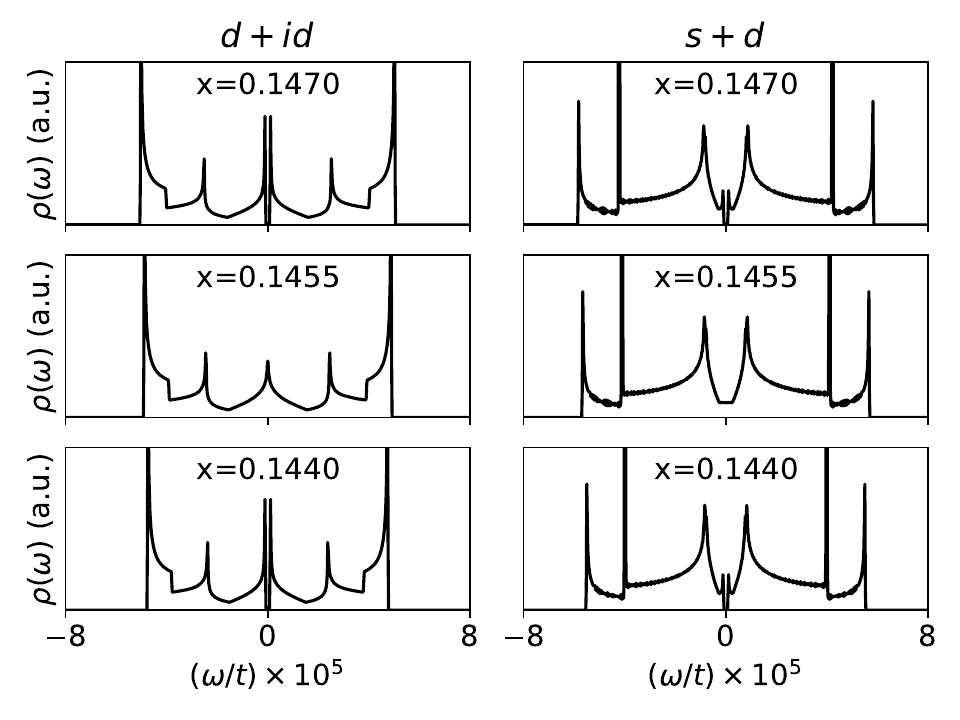}
    \caption{Low-energy density of states $\rho(\omega)$ of the $s+d$ and $d+id$ superconducting states around optimal doping $x = 0.1455$ per unit cell. Here $\omega$ is the energy of the excitation.}
    \label{fig:dos_near_optimal}
\end{figure}

The gap to the lowest-energy excitations in a multi-band unconventional superconductor is not given simply by $|\Delta_{\bm k}|$ \cite{Merino}. Therefore we explicitly calculate the momentum dependence of the superconducting gap, $\Xi_{\bm k}$, from  the eigenvalues of the mean-field Hamiltonian \cref{eq:H_mf_realspace}. That is, $\Xi_{\bm k}=E^+_{\bm k}-E^-_{\bm k}$, where $E^+_{\bm k}$ ($E^-_{\bm k}$) is the smallest positive (largest negative) eigenvalue of $H_{mf}$ at a given ${\bm k}$. This allows for resolution of the gap over the first Brillouin zone, shown in \cref{fig:gap_k} for the $s+d$ and $d+id$ superconducting states. The $s$ state is isotropically gapped. The $s+d$ and $d+id$ states are fully gapped everywhere except at optimal doping. The closing of the gap can be verified by inspection of the density of states, \cref{fig:dos_near_optimal}. Due to the flat bands in the non-interacting model, the Fermi surface covers the Brillouin zone and anywhere the gap vanishes is thus a node.

\section{Multi-orbital model}
\label{sec:multiorbsupp}
We use the Slater-Koster method \cite{SlaterKoster} to calculate hopping amplitudes between pseudo-atomic orbitals to parameterize the tight-binding Hamiltonian given by Eq.  (1). We consider $\pi$-orbitals on the ligands (L) and $d$-orbitals on the metals (M), with possible nearest-neighbor hopping between pairs of M-L, M-M, and L-L. The functions determining the relevant hoppings can be obtained from Table I of Ref.~\cite{SlaterKoster} or Table 4 of Ref.~\cite{koskinen2009}, where we treat the $\pi$-orbitals of the ligands as $p$-orbitals. 

For brevity, we illustrate below how one of the hopping amplitudes is obtained, with the hopping amplitude between other orbitals acquired in a similar manner. The hopping amplitude between a $\pi_x$ orbital and a $d_{xy}$ orbital is given by
\begin{align}
t_{x,xy} = \sqrt{3}l^2mV_{pd\sigma} +m(1-2l^2)V_{pd\pi},
\label{eq:SK}
\end{align}
which depends only on the unit displacement vector between the orbitals, $\hat{d} =(l,m,n)$, where $l, m, n$ are the $x$, $y$ and $z$ components determined from the center of the metals and ligands on the kagome-Lieb lattice; and constants, $V_{\eta \zeta \beta}$, which depend on the type of orbitals the electron is hopping between ($\eta$, $\zeta$ = $s$, $p$, $d$) and the bonding character between those orbitals ($\beta = \sigma, \pi, \delta$). The constants $V_{\eta \zeta \beta}$ are treated as free parameters that depends on the characteristics of the material of interest, and we include an on-orbital energy level, $\delta_{\mu}$, on the $d$-orbitals.

\begin{table}
\centering
\begin{tabularx}{\columnwidth}{l | 
                                >{\centering\arraybackslash}X  
                                >{\centering\arraybackslash}X 
                                >{\centering\arraybackslash}X 
                                >{\centering\arraybackslash}X 
                                >{\centering\arraybackslash}X 
                                >{\centering\arraybackslash}X 
                                >{\centering\arraybackslash}X 
                            }
    & $V_{pd\sigma}$ & $V_{pd\pi}$ & $V_{dd\sigma}$ & $V_{dd\pi}$ & $V_{dd\delta}$ & $V_{pp\sigma}$ & $V_{pp\pi}$ \\ \hline
    Even & 1 & 1 & 0.5 & 0.3 & 0.2 & 0.05 & 0.05 \\
    Odd  & 0 & 1 & 0   & -1  & 0.3 & 0.05 & 0
\end{tabularx}
\caption{Slater-Koster  energy integrals, $V_{\eta\xi\beta}$,  used in Fig. 4. 
The set of even (odd) orbitals is $\pi_x$, $\pi_y$, $d_{xy}$, $d_{x^2-y^2}$ ($\pi_z$, $d_{xz}$, $d_{yz}$).
The on-orbital energies of the $d$-orbitals are $\delta_{xz}=\delta_{yz}=0.5$, $\delta_{xy}=1.5$ and $\delta_{x^2-y^2}=4.5$, and we set the nearest-neighbor ligand-ligand hopping within hexagons and triangles equal (parameterized by $V_{pp\sigma}$ and $V_{pp\pi}$).
}
\label{tab:slater-koster-v-params}
\end{table}

The parameters that produce Fig. 4 are given in \cref{tab:slater-koster-v-params}. 
These were chosen by visual inspection to qualitatively match the band structure of Cu-BHT \cite{Cu-BHT-theo}.
To limit the search in such a large parameter space we constrained $\delta_{\mu}$ and $V_{\eta\zeta\beta}$ based on some physical considerations. The choice of the set $\{\delta_{\mu}\}$ is constrained as we require it to reproduce the expected orbital energies of the approximately  square planar inner coordination sphere of the kagome-Lieb lattice. 
$V_{\eta\zeta\beta}$ is often a decaying function of distance, such that next-nearest-neighbor hopping can be an order of magnitude smaller than nearest-neighbor hopping. However, in many MOFs M-M hopping can be comparable in strength to M-L hopping \cite{KennyE.P2021TAtP}. 
Hence, in our parameter search we chose hopping strengths $\textrm{M-L} \geq \textrm{M-M} \gg \textrm{L-L}$.

\begin{figure*}
    \centering
    \includegraphics[width=0.9\linewidth]{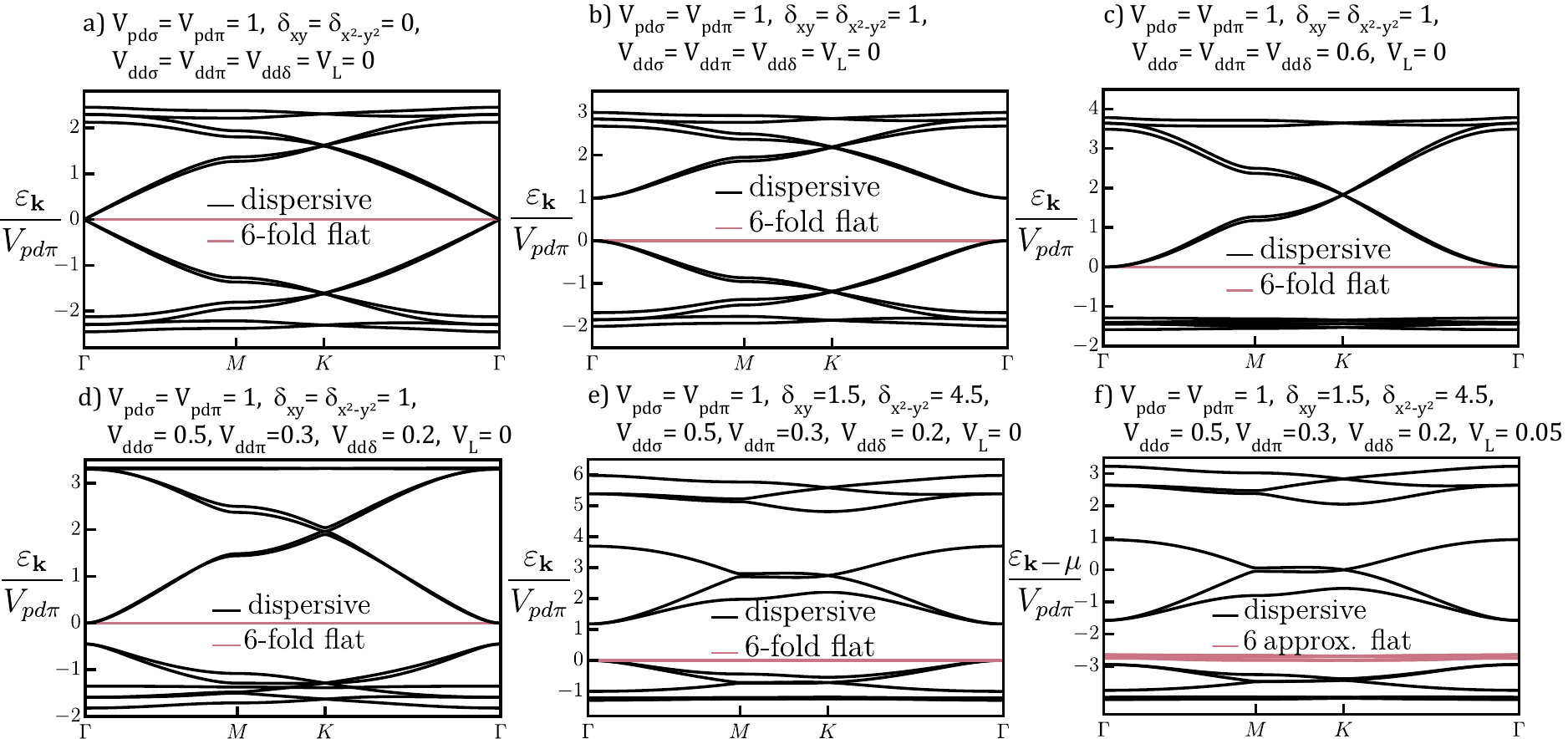}
    \caption{Evolution of the bands arising from the even orbitals as Slater-Koster parameterized M-L, M-M, and L-L hopping are varied. In (a) there is only M-L hopping, (b) includes an on-orbital energy on the $d$-orbitals, (c) introduces M-M hopping that is equal for all $d$ orbitals, (d) has unequal M-M hopping, (e) splits the on-orbital energy of the $d$-orbitals, and (f) introduces L-L hopping. (f) reproduces the even set of bands in Fig. 4, and we have shifted the energies to match the chemical potential in the band structure of Cu-BHT \cite{Cu-BHT-theo}.}
    \label{fig:multiorbitalsupp} 
\end{figure*}

In \cref{fig:multiorbitalsupp}, we show the evolution of the band structure from the simplest multi-orbital model, with nearest neighbor hopping only, to a model that closely resembles Cu-BHT. Note that the 6 nearly flat bands remain throughout. This shows that the flat bands in our model of Cu-BHT arise from Sutherland's theorem.

\newpage

%